\documentstyle[sprocl,epsfig]{article}
\def\be{\begin{equation}}
\def\ee{\end{equation}}
\def\bea{\begin{eqnarray}}
\def\eea{\end{eqnarray}}

\begin{document}
\title{\bf{WEAK FORM FACTORS OF THE NUCLEON
{\footnote{Invited talk, WEIN-98, Santa Fe, NM, June 15, 1998.\\ 
My web address : http://www.rpi.edu/\~{}mukhon\\
e.mail address : mukhon@rpi.edu}}}}
\author{ NIMAI C. MUKHOPADHYAY}
\address{Department of Physics, Applied Physics and
Astronomy\\Rensselaer Polytechnic Institute\\
Troy, New York 12180-3590, USA}
\maketitle\abstracts{
We limit ourselves basically to the $SU(2)$ flavor sector of 
the CKM matrix, as probed in processes like nuclear $\beta$-decay, 
normal and radiative muon capture and neutrino reactions. Current interests 
in this primarily stem from our aspirations to interpret 
the weak form factors of the nucleon in the framework of 
QCD. Given a standard set of known form factors, we can look also in these 
processes for possible physics signals beyond the standard model.} 

\section{Introduction}

You have all heard from Paul Langecker and others how great is the standard 
model (SM) in describing strong, electromagnetic and weak interactions. 
So what I say here will be a modest footnote to that theme. I shall pose 
the simple question : What do we know about the weak form factors of 
the nucleon and what interest does this knowledge have in the context of 
the SM? Interests of the weak form factors are not new : right 
after the discovery of the parity violation in the weak interaction, these 
interests started to develop, already quite intensely in the sixties. 
So what I say will have a familiar ring to the older members of 
our audience. Current interests in the weak form factors of nucleons 
originate in our desire to interpret them in the framework of 
quantum chromodynamics (QCD). This is a difficult subject in theoretical 
physics, as the low-$q^{2}$ QCD is highly non-perturbative and can be only 
rigorously tackled on the lattice by numerical methods that are highly 
computer-intensive. The other interest of the weak hadron form factors is to 
take a set of standard values of them and to see if familiar weak processes 
lead to any result that is beyond the scope of the SM. I shall focus here on 
the former, and refer you to reviews, for example, by Herczeg \cite{1} for 
the latter.

An outline for the remainder of this talk is as follows : In section 2, 
I shall discuss the traditional versus the QCD motivated ways of 
looking at the weak nucleon form factors. Informations 
from nucleonic processes will be reviewed in section 2, while those 
from the precesses in complex nuclei will be very briefly touched in 
section 3. I shall close with a brief summary and an outlook.

We restrict ourselves to the nucleon charged current in this talk.

\section{Traditional versus QCD-motivated ways of looking at the weak nucleon form factors}
\subsection{Some definitions}
The nucleon charged weak current has the well-known \cite{2,3} 
Lorentz structure :
\begin{eqnarray}
V_{\lambda} &=& g_{V}(q^{2}) \gamma_{\lambda} + i g_{M}(q^{2}) 
\sigma_{\lambda\nu} {q^{\nu}\over 2M} + g_{S}(q^{2})
{q_{\lambda}\over m_{l}},\\
A_{\lambda} &=& g_{A}(q^{2}) \gamma_{\lambda} \gamma_{5} + g_{P}(q^{2})
{q_{\lambda}\over m_{l}} \gamma_{\lambda} + i g_{T}(q^{2}) 
\sigma_{\lambda\nu}\gamma_{5} {q^{\nu}\over 2M}\,.
\end{eqnarray}
The notations are standard : $m_{l}$ is the charged lepton mass, $M$, the nucleon mass, $q_{\lambda}$, the momentum transfer, $n_{\lambda} - p_{\lambda}$,
$n,\,p$ being neutrons and protons. Since nucleon has an internal structure,
\begin{equation}
g_{A}/g_{V} \ne 1,\,g_{M},\, g_{S},\,g_{P} \,\,{\hbox{and}}\,\, g_{T} \ne 0,
\end{equation}
in principle. Hence the name of the last four objects  the ``induced'' nucleon form factors. From  T invariance,
\begin{equation}
Im(g_{i}) = 0.
\end{equation}
\subsection{T invariance}
This can be tested, for example, in the polarized 
neutron $(\vec{n})\, \beta$-decay \cite{4}. The phase $\phi_{AV}$ of 
$g_{A}$ can be determined by measuring the triple correlation coefficient 
$D$ from the component of $\vec{n}$ spin perpendicular to 
the decay plane in the $\beta$-decay. For T invariance,
\begin{equation}
D = 0,
\end{equation}
implying the phase
\begin{equation}
\phi_{AV} = 0^{\circ}\,\, {\hbox{or}}\,\, 180^{\circ}.
\end{equation}
The Particle Data Group (PDG) \cite{5}, in its latest compilation 
(PDG-98) reports
\begin{equation}
\phi_{AV} = 180.07 \pm 0.18\,\, degrees.
\end{equation}
We should note that there is a sign difference in $D$ between the last two 
measurements \cite{4} which were reported in the seventies. So Freedman 
and others in the audience should take a note of this and perhaps can mount 
a modern experiment to improve upon this. No real test for T violation 
in muon capture so far, though there are theoretical discussions \cite{6} 
(for $\beta$-decay theory and experiment review, see Herczeg \cite{1}) 
and an experimental proposal \cite{7} at PSI to do this. 
Deutsch \cite{7} has spoken on this subject at many conferences with a lot of 
passion. We should note that the discovery of the T-violation in the nuclear
 muon capture (NMC) will be a glimpse at the physics beyond the SM.

\subsection{Other constraints on vector form factors}
Let us continue our discussion with the vector form factors in Eq. (1). 
The hypothesis of conserved vector current, which also follows from the SM, 
implies
\begin{equation}
g_{S}(q^{2}) = 0.
\end{equation}
This is not well-tested, though some crude tests are 
available \cite{8} from the NMC.

As regards to the other vector form factors, the $SU(2)$ structure of 
the electromagnetic and weak currents allows us to 
relate the weak form factors $g_{V}(q^{2})$ and $g_{M}(q^{2})$ to 
the electromagnetic form factors of the nucleons. In the usual notation
\begin{eqnarray}
g_{V}(q^{2}) &=& F^{p}_{1}(q^{2})  -  F^{n}_{1}(q^{2}),\\
g_{M}(q^{2}) &=& F^{p}_{2}(q^{2})  -  F^{n}_{2}(q^{2}).
\end{eqnarray}
Thus, their $q^{2} \rightarrow 0$ limit and $q^{2}$ dependence are well-known;
\begin{equation}
g_{V}(0)= 1,\,\, g_{M}(0) = \kappa_{p} - \kappa_{n},
\end{equation}
where $\kappa_{p}$ and $\kappa_{n}$ are the appropriate nucleonic anomalous 
magnetic moments that are extremely well-known \cite{5}. The dipole 
dependence of these form factors is theoretically supported, 
at large $q^{2}$, by the principles of perturbative QCD \cite{9}; 
at low $q^{2}$, this is independently verified by the experiments 
on neutrino scattering \cite{10}. In this context it is interesting to 
note that the anomalous magnetic moments $\kappa_{i}$ enter in the famous 
spin-dependence sum rule, known as the Drell-Gerasimov-Hearn (DGH) sum rule 
\cite{11}, in the helicity-separated electromagnetic cross-sections 
$\sigma_{1/2}$, $\sigma_{3/2}$:
\begin{equation}
- {2\pi^{2}\alpha \kappa_{i}^{2}\over M^{2}} = \int{{d\nu\over \nu}\, 
(\sigma_{1/2} - \sigma_{3/2})},
\end{equation}
with usual symbols. Many experimental initiatives are underway to determine 
the neutron electromagnetic form factors and to test the DGH sum rule. 
Laboratories doing this include Bonn, Mainz and CEBAF, just to name three.

Continuing with the vector form factors, there are new CVC test proposals 
presented by van Schagen {\it et al.}\cite{12} and Beck {\it et al.}\cite{12} 
at this conference on the mass-12 system, wherein $\beta^{\mp}$ decays 
take place from $^{12}B$ and $^{12} N$ to $^{12} C_{g.s.}$ 
and the analogous state in $^{12} C^{*}$ also decays to the $^{12} C$ 
ground state. The corresponding reaction in the NMC is
\begin{equation}
^{12} C + \mu^{-}(1S) \longrightarrow ^{12}B_{g.s.} + \nu_{\mu}\,,
\end{equation}
a reaction well-studied theoretically \cite{13} and experimentally \cite{14}. 
It is a great experimental achievement to 
have studied the corresponding reactions by neutrino scattering by the LSND 
\cite{15} and KARMEN~\cite{15} collaborations, backed by solid 
theoretical efforts 
\cite{16} which will be summarized by Vogel \cite{17} at this conference. 
Given the excitement on neutrino oscillation at this conference, we look 
forward to more excitement from these nuclear reactions induced by neutrinos.

\subsection{Axial vector form factors}
\textbf{$2.4.1 \,g_{A}$}

\centerline{}

We now come to the nucleon axial-vector form factors. First point to make is 
the emphasis of the muon-electron universality, a cornerstone of 
the CKM theory in the SM, modulo radiative corrections \cite{2}, within which
\begin{equation}
\left(g_{A}/g_{V}\right)^{\mu} = \left(g_{A}/g_{V}\right)^{e},
\end{equation}
at $q^{2} = 0$, where $\mu$ and $e$ respectively refer to 
the NMC and $\beta$-decay. The former is extracted from the process
\begin{equation}
\mu^{-} + p \longrightarrow n + \nu_{\mu}.
\end{equation}
The latter is, of course, given by the classic reaction of 
neutron $\beta$-decay \cite{2,3,5}
\begin{equation}
n \longrightarrow p + e^{-} + \bar{\nu}_{e}.
\end{equation}
Of course, this universality test (14) can never rival the one done via the 
two-body $\beta$-decays of pions, from which one can extract the ratio 
\begin{equation}
R_{o} = \pi^{-}_{e2}/\pi^{-}_{\mu 2}\,,
\end{equation}
which is given by the kinematic expression
\begin{equation}
R_{o} = x(1-x)^{2} y^{-1} (1-y)^{-2} \,,
\end{equation}
on top of which radiative corrections must be included accurately. 
In Eq. (18), $x$ and $y$ are given by the mass ratios
\begin{equation}
x = (m_{e}/m_{\pi})^{2}\,,\,\,\, y = (m_{\mu}/m_{\pi})^{2}.
\end{equation}
The meson factories have made the measurement of the ratio $R_{o}$ 
an extremely accurate business, but we do not have time to go into it.

Unfortunately, measurements of $(g_{A}/g_{V})^{e}$ from the process 
(16) have been accurate, but not without fluctuations in the value of this 
ratio. Thus, in 1959, this ratio was known accurately enough, but quite 
different from its value as we know now \cite{5} :
\begin{equation}
(g_{A}/g_{V})^{e}_{59} = 1.17 \pm 0.02\,,\,\, 
(g_{A}/g_{V})^{e}_{98} = 1.2601 \pm 0.0025.
\end{equation}
Hence Holstein's joke \cite{18} that this fluctuation in various epochs seems 
like following the expectation of Dirac's law of large numbers!
      
In order to compare the $g_{A}/g_{V}$ from the $\mu$-capture reaction (15) to 
its value from the $\beta$-decay (16), we have to take out the modest 
$q^{2}$ dependence in the former. For this, 
we use the recent neutrino scattering experiment \cite{10}. This yields
\begin{equation}
g_{A}(q^{2})/g_{A}(0) = \left(1- {q^{2}\over M_{A}^{2}}\right)^{-2},
\end{equation}
where the $M_{A}$ is extracted from a fit to the neutrino scattering data
\begin{equation}
M_{A} = \left(0.89^{+0.09}_{-0.09}\right) \,\,{\hbox{GeV}}\,.
\end{equation}

The behavior in (21) can be understood in the framework of the extended 
Skyrmion model \cite{19}. From the world average of the singlet NMC rate in 
hydrogen \cite{2},
\begin{equation}
\Lambda_{S} = 661 \pm 47 s^{-1}\,,
\end{equation}
we can extract $(g_{A}/g_{V})^{\mu}$ at $q^{2} = 0$, using (21), (22) 
and canonical values of the other form factors, to be
\begin{equation}
(g_{A}/g_{V})^{\mu} = 1.24 \pm 0.04.
\end{equation}
Comparing (20) and (24), the $\mu - e$ universality is confirmed, 
though not at the accuracy of the ratio (17). Further improvements 
require a lot more accurate experimental efforts \cite{20} and theoretical 
studies on the radiative corrections \cite{21} in these processes.
\\
\\
\textbf{2.4.1.1 Axial vector coupling constant as a benchmark :}
\\

Though not the principal subject of this talk, it is worthwhile to refer to 
the important role $g_{A}$ plays in various areas in physics. We shall 
not delve here its importance in astrophysics, though that would 
be a talk in itself.\\
\\
\textbf{2.4.1.1.1 Bjorken (Bj) sum rule}\\

The Bj sum rule \cite{22} is a QCD benchmark in which $g_{A}$ is connected to 
the physics of deep inelastic scattering (DIS) in nucleon.

To summarize this, we start with polarized 
structure function $g_{1}$ and $g_{2}$, which can be written in terms of 
the polarization observables
\begin{equation}
d\sigma^{\uparrow \downarrow} - d\sigma^{\uparrow \uparrow} = 
a g_{1}(x, Q^{2}) + b g_{2}(x, Q^{2})\,,
\end{equation}
where $x$ is the well-known Bj $x$, and $a, b$ are kinematic quantities, 
$b$ being small relative to $a$. Defining
\begin{equation}
\Gamma^{i}_{1} = \int^{1}_{0} g_{1}^{i}(x)\, dx \,, \,\,i = p,n\,,
\end{equation}
we have the famous Bj sum rule, modulo QCD radiative corrections which are 
well-investigated \cite{23} theoretically. Many experiments \cite{24} 
at CERN (EMC, SMC), SLAC (E142, E143, E154, E155) and HERA (HERMES) 
have determined $\Gamma^{i}_{1}$, though there are always issues of 
extrapolation. For example, a typical EMC number for $\Gamma^{p}_{1}$ is 
\begin{equation}
\Gamma^{p}_{1} = 0.126 \pm 0.01\, (stat) \pm 0.015\, (syst).
\end{equation}
In Fig.1 of Mulders and Sloan \cite{23}, it is demonstrated 
how the Bj sum rule is tested by taking into 
QCD corrections to the sum rule to ever higher orders \cite{25}. Thus, this 
sum rule has become a tool of testing QCD to a high accuracy.
\\
\\
\textbf{2.4.1.1.2 Connection to resonance physics : Adler-Weissberger sum rule}\\

It is interesting that chiral current algebra allows one to 
write the famous Alder-Weissberger sum rule \cite{26} in terms of 
strong interaction observables. In the usual notation, this sum rule 
relates integral over $\pi^{\pm} p$ cross-sections for zero mass pions :
\begin{equation}
\left[g_{A}(0)\right]^{-2} = 1 + const \int^{\infty}_{\nu_{t}} 
{d\nu^{\prime}\over \nu^{\prime}} \left[ \sigma^{0}(\pi^{-}p,\nu^{\prime})
-  \sigma^{0}(\pi^{+}p,\nu^{\prime}) \right].
\end{equation}
This is an example of the low energy theorem or soft-pion theorem 
at work. These cross-sections are, of course, resonance dominated, 
thus bringing a connection between weak and strong interaction dynamics, 
Adler used his technique of off-mass shell corrections to 
get the magnitude of $g_{A}(0)$
\begin{equation}
g_{A}(0)|_{\hbox{Adler}} = 1.24 \pm 0.03\,,
\end{equation}
which is in fine agreement with the latest experimental result quoted earlier.
\\
\\
\textbf{2.4.1.1.3 Connection to strong interaction 
dynamics in various QCD-inspired models}\\

In quark model \cite{27} and Skyrme model \cite{19}, the value of $g_{A}$ is 
not entirely trivial. Indeed, in quark model, one must take into account 
relativistic effects in some fashion, while, in the Skyrme model, 
the dynamics goes astray without a lot of finetuning of the so-called 
anomalous sector~\cite{19}. A reasonable description of $g_{A}$ has been 
found in the lattice gauge calculation within its systematic error \cite{28}. 
In the QCD sum rule approach, $g_{A}$ has recently been computed by Ioffe 
\cite{29} to be
\begin{equation}
\left(g_{A}\right)_{QCDSR} = 1.37 \pm 0.10.
\end{equation}
\\
\\
\textbf{2.4.2 $g_{P}(q^{2})$}\\

Though the discussion of $g_{P}(q^{2})$ goes back to the old works of 
Wolfenstein (who is fortunately in the audience), Goldberger and Treiman 
(GT)\cite{30}, the work of Wolfenstein continues to simulate current theoretical 
discussions both in the lattice gauge theoretic context \cite{28} 
and in the chiral perturbation theoretic ($\chi$PT) framework \cite{30}. It 
continues to be the primary focus of the NMC in hydrogen \cite{2} and of 
the $(e, e^{\prime} \pi)$ reaction \cite{32} near threshold. 
Either via once-subtracted dispersion relation of 
Wolfenstein or via chiral Ward identity \cite{30}, we can write
\begin{equation}
g_{P}(q^{2}) = {2 m_{\mu} g_{\pi NN} F_{\pi} \over m_{\pi}^{2} - q^{2}}
\,- \, {1\over 2} g_{A}(0) m_{\mu} M r_{a}^{2}\,,
\end{equation}
with $g_{\pi NN}$ and $F_{\pi}$ being the pion-nucleon coupling constant 
and pion decay constant respectively, $r_{a}$ being the nucleon axial vector 
radius. We recognize the first term on the right-hand side of Eq. (31) as 
the old Nambu-GT term, the second term being the Wolfenstein correction. For 
NMC in hydrogen, $q^{2} = q^{2}_{0} = - 0.88 m_{\mu}^{2}$, and 
\begin{equation} 
g_{P}(q^{2}_{0}) = 8.44 \pm 0.23.
\end{equation}
Recently Bernard {\it et al.}\cite{30} have made an important observation that 
the relation (31) is chirally protected and thus its experimental test is 
an important physics goal.\\
\\
\textbf{2.4.2.1 How good is the GT relation?}\\

Eq(31) can be approximately written as the GT relation (GTR) :
\begin{equation} 
F_{\pi} g_{\pi NN}(0) = \sqrt{2} M g_{A}(0)\,,
\end{equation}
where $\sqrt{2}$ is by convention. One can thus write a discrepancy function 
for the GTR after Hemmert {\it et al.}\cite{10} :
\begin{equation} 
\Delta_{\pi} = 1 - {\sqrt{2} M g_{A}(0) \over F_{\pi} g_{\pi NN}(m_{\pi}^{2})}.
\end{equation}
We can rewrite (34) as 
\begin{equation} 
\Delta_{\pi} = 1 - {g_{\pi NN}(0)\over g_{\pi NN}(m_{\pi}^{2})}.
\end{equation}
Already in 1964, Bjorken and Drell declared that the GTR ``agrees 
with experiment to better than 10\%''. Theoretically, the expectation of 
the agreement of the GTR with experiment is much higher. For 
example, Holstein reports \cite{18} two numbers. Using the linear 
$\sigma$-model, $\Delta_{\pi}$ is estimated to be 
\begin{equation}
\Delta_{\pi}^{th} \sim 0.02\,.
\end{equation}
Using the Dashen-Weinstein $SU(3)$-theoretic theorem
\begin{equation}
\Delta_{\pi}^{th}  =   0.028\,.
\end{equation}
The hypothesis of partial conservation of axial current (PCAC) gives
\begin{equation} 
g_{\pi NN}(0) \sim  g_{\pi NN}(m_{\pi}^{2})\,.
\end{equation}
This gives $\Delta_{\pi}$ to be identically zero.

We also note here that Hemmert {\it et al.}~\cite{10} have found a similar 
relation in the excitation of the $\Delta(1232)$, originally discussed 
by Adler \cite{31}, an analogue of the GTR for resonance excitation, with 
similar expectation of theoretical accuracy. This relation has raised 
recent theoretical~\cite{33} and experimental~\cite{33} interests, latter at CEBAF.
\\
\\
\\
\textbf{2.4.2.2 Status of $g_{\pi N N}$}\\

Ericson has recently \cite{34} raised 
the question, which must be in the mind of a lot of us, as to 
how our world would be if $g_{\pi N N}$ changed significantly from what it is 
known to be! Of course, such questions can be rhetorically raised about 
any physical constant. In this context, we should note that 
the VPI and other groups \cite{35} have recently differed quite a bit as to 
this quantity. To take one example, the value of $g_{\pi N N}$ equal to 
13.4 would make  the $\Delta_{\pi}$ to be 4.1\%, while its value of 
13.05 would yield $\Delta_{\pi}$ to be 1.5\%. 
This is a point much emphasized by Holstein 
recently. We may add that the PSI proposal PSI-98.01 by Oades 
{\it et al.} will measure the 1S width of 
the pionic hydrogen $\Gamma_{1S}(\pi^{-}p)$ 
at an accuracy of 1\%, which, via the Gell-Mann - Oehme sum rule, would 
yield a precise estimate of $g_{\pi N N}$ better than what is available.
Here is an example of {\it strong} interaction 
quantity influencing mightily our knowledge of the nucleon 
{\it weak} form factors!
\\
\\
\textbf{2.4.2.3 What do we know about $g_{P}$ from the NMC?}\\

As you all know, the role of $g_{P}$ is negligible 
in the $\beta$-decay, but not in the NMC. Let me recall the argument for 
the benefit of students, if any, in the audience. 
The $\beta$-decay matrix element $<M E>_{\beta}$ is proportional to 
the following expansion :
\begin{equation}
<M E>_{\beta}\, \sim 1 + 6.7 {m_{e} \over m_{\mu}} {m_{e} \over 2M} 
\,\,\sim\, 1 + O (10^{-5})\,,
\end{equation}
the second term being the reduction of the $\gamma_{5}$ operator in the non-relativistic sense. Similarly, for the NMC,
\begin{equation}
<M E>_{\mu}\, \sim 1 + 6.7 {m_{\mu}\over 2M}\,, 
\end{equation}
the second term on the right-hand side being considerably larger 
than the corresponding term in the $\beta$-case and being of 
the same order as the weak magnetism term $g_{M}$. 
Thus, using our present experimental knowledge of $\Lambda_{S}$ 
from the NMC in hydrogen, we get, fixing other form factors at their canonical values, 
the value of $g_{P}$ to be
\begin{equation}   
g_{P}(q^{2}_{0}) = 8.7 \pm 2.9\,,
\end{equation}
compared with the PCAC ($\chi$PT), value \cite{30}
\begin{equation}
g_{P}(q^{2}_{0})|_{PCAC} = 8.44 \pm 0.23\,, 
\end{equation}
a fine agreement, though there is a lot of room for improvement from better
precision of the $\Lambda_{S}$ measurement to be attempted in future~\cite{36},
as reported in a new PSI proposal, presented in a poster session at the
WEIN-98
by C. Petitjean and P. Kammel~\cite{36}. This proposal wants to 
reach a precision of $\Lambda _S$ to 1 \%. It would be great if this goal is 
reached in future.

Does it mean we are in good shape here? Not quite, as the folks from TRIUMF
have come with a disturbing result in the radiative muon capture (RMC) discussed
below~\cite{37}.\\
\\
\textbf{2.4.2.4 What is the latest from the radiative muon capture?}\\

As we know from many discussions in the literature, a very powerful way to
determine $g_p$ is from the RMC in hydrogen:
\begin{equation}
\mu ^{-} + p \quad \longrightarrow \quad n + \nu _{\mu} + \gamma
\end{equation}
wherein the high-energy photon is particularly sensitive to $g_p$. As we all
know, the RMC is much weaker than the ordinary NMC and we are talking about a
small part of the photon spectrum. Thus, we are talking about a very
difficult
experiment. The TRIUMF team has really pulled together a great {\it tour de
force}~\cite{37}
by using high muon flux on a very pure liquid hydrogen target, avoiding the
processes such as:
\begin{eqnarray*}
\pi ^{-}p & \longrightarrow  & \pi ^{o}n\,, {\hspace{3.25in}} (43a)\\
\pi ^{o}  & \longrightarrow  & \gamma \gamma n \,(55-83\,{\hbox{MeV}}\, 
\gamma\, {\hbox{'s}})\,,{\hspace{2.15in}} (43b)\\
\pi ^{-}p & \longrightarrow  & \gamma n\, (k \sim 129\, {\hbox{MeV}})\,.
{\hspace{2.45in}} (43c)
\end{eqnarray*}
The outcome of this great experiment is a puzzle to the theorists. Expressing
it in terms of an extracted $g_p$, we get from this experiment~\cite{37}
\begin{equation}
g_p(-0.88\,m_{\mu }^{2})  = (9.8\pm 0.7\pm 0.3)\,g_A(0)\,,
\end{equation}
which is 1.46 times the value expected from the $\chi$PT considerations.

This experiment has generated a cottage industry of higher order $\chi$PT
calculations~\cite{38}, using the heavy baryon $\chi$PT (HB$\chi$PT). 
These use the Ecker-Moj$\check{z}$i$\check{s}$
Lagrangian using the development upto order O($p^3$). However, these higher
order corrections do not alter significantly the leading order result. One
point of caution: building $\Delta(1232) $ effects in the $\chi$PT is 
technically
difficult~\cite{39} and I cannot judge at this time how good these 
calculations are,
looked from this point of view.

Assuming that the discrepancy between the $\chi$PT theoretical expectation and
(44)
is not due to any higher order correction in the $\chi$PT (or PCAC), what are the
possible ways out? Let me speculate on two prospects. First is the
possibility
that the above discrepancy is due to our lack of complete understanding of
the
complex molecular physics problem in the $p\mu ^{-}p$ system involved. This
means more precise experiment on the ortho($L=1$) and para($L=0$) transition
rates. Direct measurement on this is underway at TRIUMF (expt.766-96). Second
possibility has been discussed by Opat long ago. He mentioned multiple pion
exchange mechanism that distinguishes ordinary NMC and RMC. Is it possible
that
this mechanism is doing the trick alone? One needs another theoretical
study. Of course, there can always be something that we have not
considered above!\\
\\
\textbf {2.4.2.5 The $\pi $NN and pseudoscalar form factors from the lattice
QCD}\\

Thanks to the heroic efforts 
by K.-F.Liu and collaborators~\cite{28}, we have new
insights on the pion-nucleon and the pseudoscalar couplings from a quenched
lattice calculation on a $16^3\times 24$ lattice, implementing thereon the
appropriate Chiral Ward identity.\\ 
%




 The lattice measurement of the $g_{\pi NN}(q^2)$ cannot distinguish between
mono-pole and dipole behavior,except at very low $q^2$. A quantity $h_A(q^2)$,
related to our pseudoscalar form factor, nicely probes the one-pion tail, the
classic Nambu behavior~\cite{30}, and agrees very well with the experiment 
of Choi {\it et al.}
on the electroproduction of pions~\cite{32}. Thus, the "lattice" and "real"
measurements
match very well! An excellent demonstration of QCD in its non-perturbative
domain! New technique~\cite{41} on the handling of chiral symmetry breaking of fermions
on the lattice adds a lot of excitement to the improvement of the lattice
result in near future. So please stay tuned!\\
\\
\textbf {2.4.3 "Second-class" currents?}\\

Weinberg~\cite{42} , in his classic 1958 paper, defined two classes of weak hadron
currents. The "first class" ones behave under the G-parity transformation
G($G=Ce^{i\pi \tau _2}$, C, charge conjugation, $\tau _2$, the appropriate
isospin generator) as follows:
\begin{equation}
GV_{\lambda }G^{-1} = V_{\lambda },   GA_{\lambda }G^{-1} = -A_{\lambda }\,.
\end{equation}
These which do not behave like (45) are "second-class" and Weinberg surmised
that these should be absent or highly suppressed. This important conjecture is
also supported by the renormalizability of the gauge theories. The absence of
the second-class currents(SCC) implies
\begin{equation}
g_S(q^2) = 0\,,
\end{equation}
also required by the conserved vector currents(CVC), and
\begin{equation}
g_T(q^2) = 0\,.
\end{equation}
 There are related theorems by Cabibbo~\cite{43} . Search for the SCC's in the
$\beta $-decay was systematically started by Wilkinson~\cite{44}  (fortunately for us, in
the audience) and soon extended to the NMC~\cite{45} . This search has so far been
consistent  with its absence, but the subject is still quite healthy. Indeed,
the WEIN-98 has seen two contributions, one from van Schagen {\it et al.}~\cite{12}  on the
possible test of the CVC/SCC in the $A=8$ isomultiplet and the other is by
Minanisono {\it et al.}~\cite{46}  on the correlations in the spin-aligned mirror pairs in the
$A=12$ system. Let us wait for these studies for better limits than what we
have now. The latter work claims that
\begin{eqnarray}
2M\, \frac{g_T}{g_S} = 0.22 \pm 0.05 \pm 0.15(syst)\nonumber \\
                                    \pm 0.05(th)\,,
\end{eqnarray}
in comparison with the expectation of the QCD sum rule~\cite{47}
\begin{equation}
2 M\,{g_{T}\over g_{A}} = 0.0152 \pm 0.0053\,,
\end{equation}
where the G-violation here is proportional to the mass difference of quarks
$m_{u }-m_d$ ; these masses also figure in the isospin violation in the
pion-nucleon scattering~\cite{48}.

 Muon capture experiments are consistent with the absence of the SCC. Two
latest limits are already quite old. First is from Holstein~\cite{49},
\begin{equation}
g_S(0) = -0.4 \pm 2.3\,,
\end{equation}
belonging to our survey of the vector form factors earlier, second is from
Morita~\cite{50},
\begin{equation}
g_T(q_0^V) = -0.06 \pm 0.49\,.
\end{equation}\\

\section{Nuclear muon capture: a brief excursion to a long and old subject
which is still very interesting}

  This is a vast subject that has been reviewed repeatedly in
literature($e.g.^2$). I cannot do justice to the vast subject accumulated
over
many years here. Let me instead pick up one process in which an enormous
progress has been made at PSI. The PSI experiment has measured the
statistical
capture rate of the muon capture in $^3He$ from the 1S atomic state to an
unprecedented accuracy. The reaction of interest is
\begin{equation}
\mu ^- + ^3He \longrightarrow \nu _{\mu } + ^3H\,.
\end{equation}
The measured rate is~\cite{51}
\begin{equation}
\Lambda _{stat}^{exp} = 1496 \pm 3(stat) \pm 3(sys)\, s^{-1}\,.
\end{equation}
The theory has been treated most recently in a number of works, most
thoroughly
by Congleton and collaborators~\cite{52}. 
Congleton and Truhlik~\cite{52} find a 15\% meson
exchange current effect, thus a significant contribution and a proof positive
on the role of the two-body currents in this clean nuclear reaction(52). They
obtain a theoretical rate~\cite{52}
\begin{equation}
\Lambda _{stat}^{th} = 1497 \pm 21\, s^{-1}\,,
\end{equation}
in excellent agreement with (54).\\

 The experimental result (54) is so precise that one can explore it for a
variety of physics effects, including an explosion of the physics beyond the
standard model, as Govaerts~\cite{53} has recently attempted. We refer to his papers
for
further details. Deutsch~\cite{53}, Herczeg~\cite{53} and Mohapatra~\cite{53} have inspired these
investigations
by emphasizing repeatedly the importance of the NMC as a window on the
physics
beyond the standard model.\\

 The reaction (52) has been used by Junker and this author~\cite{54} in the context of
the nuclear PCAC and GTR. Following the method of Wolfenstein~\cite{30}, one can write a
dispersion relation:
\begin{eqnarray}
D(t)= [2MF_A + \frac{t}{m}F_p]\nonumber \\
    = \frac{-\sqrt{2}F_{\pi }m_{\pi }^2G(t)}{t-m_{\pi }^2} 
- \frac{1}{\pi} \int_{a}^{\infty} dt' \frac{Im(D(t))}{t' - t}\,,
\end{eqnarray}
where "a" represents thresholds for various nuclear anomalous cuts (for $d +
n$
and $n + p + n$ channels for the t breakup). Here $F_A$,$F_P$ are the nuclear
analogue of the nucleon axial vector form factors. From this master equation,
one gets an expression for $G^{eff}$, the pion-nuclear coupling constant,
yielding:
\begin{equation}
G^{eff}(m_{\pi }^2) = 45.8 \pm  2.4\,,
\end{equation}
compared with the value obtained from the pion-nuclear scattering:
\begin{equation}
G_{st}^{eff}(m_{\pi }^2) = (38 \pm  16 )\quad to \quad (57 \pm  13)\,.
\end{equation}
{\it Thus, weak interaction in this context is more powerful than the strong
interaction.}

 There are also recent~\cite{53} limits available on the second class current
couplings from this reaction.

 The asymmetry observable in the polarized muon capture are also very
interesting~\cite{55}.

 For brevity, we omit here the important subject of the effective weak hadron
coupling constants~\cite{56} in nuclei. Suffice to say that there was a lot of interest
in the parallel sessions at this meeting on this topic, as well as many La
Fonda hallway discussions on the role of the NMC as a means to explore
physics
beyond the standard model. Recent works, besides Govaerts, by Barabanov~\cite{57},
Ciechanowicz and Popov~\cite{58}, Herczeg and Mukhopadhyay~\cite{59}, Missimer {\it et al.}~\cite{60}, should be referred for further information. 

\section{Summary}
Let me restate the main points of this talk:
\begin{itemize}
\item The $\beta $-decay and the NMC(RMC) are major physics sources for
    examining old and new issues on weak nucleon form factors, particularly
in
    the context of recent QCD interest.
\item  Radiative muon capture experiment at TRIUMF poses a challenge to our
    understanding of the application of the PCAC/$\chi$PT to this reaction.
\item The $^3He \longrightarrow ^3H$ reaction is now a new gold standard in
    the field. It contains lots of  valuable information.
\item The NMC looks promising in the context of physics beyond the standard
    model.
\end{itemize}
High stopping rate of muons, polarized nuclear target and other
    experimental innovation should make this subject one of continuing
    interest.

\section*{Acknowledgments}
I pay special thanks to four veterans of
this subject in the audience: J. Deutsch, M. Morita, D.H. Wilkinson 
and L. Wolfenstein. I have
been deeply inspired by these great people.

I sincerely thank D. D. Armstrong, J. Deutsch, P. Herczeg, C. Hoffmann, 
B. Holstein and M. Morita for many helpful comments and suggestions. This 
work has been partially supported by the U.S. Department of Energy. I am 
writing these lecture notes under the shadow of a grave illness; that I can 
write them at all seems to be a very pleasurable experience. I apologize for 
any shortcomings, particularly in not being able to cover fully the recent 
literature. I ask for forgiveness for readers and authors of the relevant 
papers not cited. A very special word of thanks to my RPI colleagues 
N. Mathur and R.M. Davidson, the former for his great effort to put the paper 
in LaTeX and the latter for a proof reading. A word of deep gratitude for 
M. and P. Herczeg and S. Mukhopadhyay for many words of encouragement in these 
difficult times.

I hope to see you at a future WEIN conference with better health.

\end{document}